\input harvmac
\newcount\figno
\figno=0
\def\N{{\cal N}}
\def\fig#1#2#3{
\par\begingroup\parindent=0pt\leftskip=1cm\rightskip=1cm\parindent=0pt
\global\advance\figno by 1
\midinsert
\epsfxsize=#3
\centerline{\epsfbox{#2}}
\vskip 12pt
{\bf Fig. \the\figno:} #1\par
\endinsert\endgroup\par
}
\def\figlabel#1{\xdef#1{\the\figno}}
\def\encadremath#1{\vbox{\hrule\hbox{\vrule\kern8pt\vbox{\kern8pt
\hbox{$\displaystyle #1$}\kern8pt}
\kern8pt\vrule}\hrule}}
\def\underarrow#1{\vbox{\ialign{##\crcr$\hfil\displaystyle
 {#1}\hfil$\crcr\noalign{\kern1pt\nointerlineskip}$\longrightarrow$\crcr}}}
%
\overfullrule=0pt

%
\def\tilde{\widetilde}

\def\Z{{\bf Z}}
\def\T{{\bf T}}
\def\S{{\bf S}}
\def\R{{\bf R}}

\font\zfont = cmss10 

\def\bigone{\hbox{1\kern -.23em {\rm l}}}
\def\ZZ{\hbox{\zfont Z\kern-.4emZ}}

\Title{hep-th/9807205, IASSNS-HEP-98/66, RU-98-34}
{\vbox{\centerline{Anti-de Sitter Space And The Center Of The Gauge Group}
}}
\smallskip
\centerline{Ofer Aharony}
\smallskip
\centerline{\it Department of Physics and Astronomy}
\centerline{\it Rutgers University, Piscataway NJ 08855, USA}
\smallskip
\centerline{Edward Witten}
\smallskip
\centerline{\it School of Natural Sciences, Institute for Advanced Study}
\centerline{\it Olden Lane, Princeton, NJ 08540, USA}
\bigskip

\medskip

\noindent

Upon compactification on a circle, $SU(N)$ gauge theory with all
fields in the adjoint representation acquires a $\Z_N$ global symmetry
because the center of the gauge group is $\Z_N$. For $\N=4$ super
Yang-Mills theory, we show how this $\Z_N$ ``topological symmetry''
arises in the context of the AdS/CFT correspondence, and why the
symmetry group is $\Z_N$ rather than $U(1)$. This provides a test of
the AdS/CFT correspondence for finite $N$. If the theory is formulated
on ${\bf R}^3\times\S^1$ with anti-periodic boundary conditions for
fermions around the $\S^1$, the topological symmetry is spontaneously
broken; we show that the domain walls are $D$-strings, and hence that
flux tubes associated with magnetic confinement can end on the domain
walls associated with the topological symmetry. For the $(0,2)$
$A_{N-1}$ superconformal field theory in six dimensions, we
demonstrate an analogous phenomenon: a $\Z_N$ global symmetry group
arises if this theory is compactified on a Riemann surface.  In this
case, the domain walls are $M$-theory membranes.

\Date{July, 1998}
\newsec{Introduction}

According to a by now celebrated conjecture \ref\malda{J. Maldacena,
``The Large $N$ Limit Of Superconformal FIeld Theories And Supergravity,''
hep-th/9711200.},
${\cal N}=4$ super Yang-Mills theory with $SU(N)$ gauge group on $\S^4$
is equivalent to Type IIB superstring theory on $AdS_5\times \S^5$,
with $N$ units of five-form flux on $\S^5$, and with the complex
coupling constant $\tau_{IIB}$ of Type IIB string theory identified
with the Yang-Mills
coupling constant $\tau_{YM} = {\theta \over {2\pi}} + {{4\pi
i} \over g_{YM}^2}$.  Correlation functions
of the gauge theory are expressed\nref\kleb{S. Gubser, 
I. R. Klebanov, and A. M. Polyakov, 
``Gauge Theory Correlators From Noncritical String Theory,'' hep-th/9802109.}
\nref\witten{E. Witten, ``Anti-de Sitter Space And Holography,'' 
hep-th/9802150.}
\refs{\kleb,\witten} in terms of the dependence of the Type IIB string theory
on boundary conditions.

The manifold $\S^4$ (which can be viewed as the conformal
compactification of $\R^4$) enters because it is the boundary, in a
conformal sense, of $AdS_5$; to study the ${\cal N}=4$ theory on
another four-manifold $M$, one would replace $AdS_5$ by a negatively
curved Einstein manifold of boundary $M$.  An interesting example is
to take $M_4=M_3\times \S^1$, with $M_3$ a three-manifold. If one uses
antiperiodic boundary conditions for fermions in the $\S^1$ direction,
then quantum field theory on $M_4$ is related to a thermal ensemble:
it computes $\Tr\,e^{-\beta H}$ with $H$ the Hamiltonian in
quantization on $M_3$.  In the present paper, we concentrate on the
cases that $M_3$ is $\S^3$ or $\R^3$.

The ${\cal N}=4$ super Yang-Mills theory has all fields in the adjoint
representation of $SU(N)$. So, if this theory is formulated on
$M_3\times \S^1$, it develops a $\Z_N$ global symmetry that has the following
origin.  Let $\tilde G_0$ be the group of  $SU(N)$
gauge transformations on $M_3\times \S^1$, and let $\tilde G$ be the
group of $SU(N)/\Z_N$ gauge transformations
on the same manifold. $\tilde G$ includes gauge transformations by
gauge functions $g:M_3\times \S^1\to SU(N)/\Z_N$ which, if lifted to $SU(N)$,
would be multiplied in going around the $\S^1$ by an element of the center
of $SU(N)$. In quantizing $SU(N)$ gauge theory
on $M_3\times \S^1$,
one divides by $\tilde G_0$, and the quotient
\eqn\thequo{\Gamma=\tilde G/\tilde G_0}
is observed as a global symmetry.  $\Gamma$ is isomorphic to the center of
$SU(N)$, which in turn is naturally identified with the group of $N^{th}$
roots of unity.  We will call any symmetry such as
$\Gamma$ which depends on the topology of spacetime    a topological 
symmetry\foot{The full topological symmetry group of ${\cal N}=4$
$SU(N)$ super
Yang-Mills theory compactified on a circle is actually
$\Z_N\times \Z_N$, as discussed at the end of section 2.}.

An order parameter for the $\Z_N$ global symmetry is a ``temporal 
Wilson line.''
Let $P$ be a point in $M_3$, let $C_P=P\times \S^1$, and let
\eqn\jicco{W(P)=\Tr\,P\exp\oint_{C_P}A,}
with the trace taken in the fundamental representation of $SU(N)$.  A
gauge transformation which in wrapping around the circle is multiplied
by an $N^{th}$ root of unity $\zeta$ multiplies $W(P)$ by $\zeta$.
Therefore, the expectation value of $W(P)$ can serve as an order
parameter for the $\Z_N$ symmetry; if $\Gamma$ is unbroken,
expectation values
\eqn\umcob{\langle W(P_1)W(P_2)\dots W(P_k)\rangle}
vanish unless $k$ is divisible by $N$.

We have so far assumed that the gauge group is $SU(N)$ rather than
$U(N)$; for $U(N)$ gauge theory, the symmetry group is $U(1)$ rather than
$\Z_N$.  There are indeed strong arguments which show that Type IIB on
$AdS_5\times \S^5$ is related to $SU(N)$ rather than $U(N)$ gauge theory
\witten.  But there is also a puzzle \ref\wittentwo{E. Witten, ``Anti-de
Sitter Space, Thermal Phase Transition, And Confinement In Gauge Theories,''
hep-th/9803131.},
since a study of a high temperature phase on $\S^3\times \S^1$ appeared
to show a $U(1)$ topological symmetry group.  We resolve this question in
section 2.  We show that on $\S^3$ (and similarly on
 any three-manifold $M_3$), the
topological symmetry group at either high temperatures or low temperatures
is $\Z_N$, not $U(1)$.  However, to show this at high temperatures requires
a novelty involving a certain spacetime $\theta$ angle.  

Since our question probes the fact that the center of the gauge group is
$\Z_N$ rather than $U(1)$, it is naturally
 closely related to the existence of a baryon vertex 
(the possibility of forming a gauge-invariant combination of $N$
external
quarks), which probes the same issue.  There have been two related but
somewhat different proposals concerning the baryon vertex: it has
been constructed from a wrapped fivebrane \ref\wittenthree{E. Witten,
``Baryons And Branes In Anti-de Sitter Space,'' hep-th/9805112.}, or 
connected with
 the existence of a certain low energy coupling $B_{NS}\wedge H_{RR}
\wedge G_5$
\ref\go{D. J. Gross and H. Ooguri, ``Aspects Of Large $N$ Gauge Dynamics
As Seen By String Theory,'' hep-th/9805129.}.
As we will see, each of these facets of the baryon vertex is helpful
in understanding the topological symmetry; the first is more directly
relevant at low temperatures, and the second at high temperatures.

The identification of the topological symmetry group is one of a few
direct tests of the conjecture of \malda\ for finite values of
$N$. Many of the direct tests of the conjecture up to now 
 test the conjecture only in the limit $N \to \infty$.

We also consider the case that $M_3$ is decompactified to $\R^3$.
Thus, in this case we are studying the ${\cal N}=4$ theory on
$\R^3\times\S^1$.  In this infinite volume situation, the topological
symmetry group is spontaneously broken, as already shown from the Anti-de
Sitter point of view   in \wittentwo.
Since the topological symmetry is a discrete symmetry, its breaking
results in the existence of domain walls.  One might think that the domain
walls would be solitonic objects in the large $N$ limit, with tensions
of order $N^2$ (the large $N$ effective theory of 
glueballs has an effective action proportional to $N^2$, and all solitonic
objects would have energy or energy density of order $N^2$). 
However, it turns out that the domain walls are in fact
$D$-strings, with tensions of order $N$.  Such behavior
is by now familiar from various examples, some of them
treated in \wittenthree, where it has turned out that various particles,
strings, or domain walls of large $N$ gauge theory, which one might
think would be solitonic objects of the large $N$ effective theory,
are better understood as $D$-branes.

In section 3, we move on to discuss the $(0,2)$ theory in six
dimensions.  The $A_{N-1}$ version of this theory is constructed, as
in \malda, via its conjectured dual which is the compactification of
$M$-theory on $AdS_7\times \S^4$, with $N$ units of four-form flux on
$\S^4$.  If this theory is compactified on a circle, it looks at low
energies like $SU(N)$ Yang-Mills theory, and hence may be expected, if
further compactified on a second circle, to exhibit a $\Z_N$
topological symmetry.  Thus, if the $(0,2)$ theory is compactified on
$\S^1\times \S^1 =\T^2$, one should expect a $\Z_N$ global symmetry.
In section 3, we establish a stronger result: we show that if the
$(0,2)$ theory is compactified on a Riemann surface $F$ of any genus,
not necessarily $\T^2$, it develops a $\Z_N$ symmetry.  This assertion
cannot at present be tested independently in any obvious way. It
seems like an interesting probe of the inner nature of the still
rather mysterious $(0,2)$ theory.

\newsec{$AdS$ Space And The Topological Symmetry Of Yang-Mills Theory}

\def\N{{\cal N}}
\subsec{Behavior On Compact Spacetime}

We will mainly consider $\N=4$ super Yang-Mills on $\S^3\times \S^1$.
This manifold is  \ref\hp{S. Hawking and D. Page, ``Thermodynamics Of
Black Holes in Anti-de Sitter Space,'' Commun. Math. Phys. {\bf 87}
(1983) 577.} the boundary 
of two different
negatively curved Einstein manifolds $X_1$ and $X_2$:

(1)  $X_1$ is
 a quotient of $AdS_5$ and has topology $D_4\times \S^1$,
where $D_4$ is a four-ball of boundary $\S^3$.

(2) $X_2$ is the Euclidean  Schwarzschild black hole solution of
Anti-de Sitter space; its  topology is $\S^3\times D_2$, where $D_2$ is a 
disc or two-ball of boundary $\S^1$.

$\N=4$ super Yang-Mills on $\S^3$ at low or high temperatures,
respectively, is described (for large $N$ and $g_{YM}^2 N$) by Type
IIB superstrings on $X_1\times \S^5$ or $X_2\times \S^5$
\wittentwo.

\nref\newmalda{J. Maldacena, ``Wilson Loops In Large $N$ Field Theories,''
Phys. Rev. Lett. {\bf 80} (1998) 4859, hep-th/9803002.}
\nref\reyyee{S.-J. Rey and J. Yee, ``Macroscopic Strings As Heavy Quarks In
Large $N$ Gauge Theories And Anti-de Sitter Supergravity,'' hep-th/9803001.}
Now we consider, as explained in the introduction, the problem of computing an
expectation value of a product of $k$ temporal Wilson lines:
\eqn\umpo{\langle W(P_1)\dots W(P_k)\rangle.}
The $P_i$ are points in $\S^3$, and the $i^{th}$ Wilson line is wrapped
on $C_i=P_i\times \S^1$.
According to \refs{\newmalda,\reyyee}, this expectation value equals a string
theory partition function in which one sums over configurations in which
the $C_i$ are boundaries of elementary string worldsheets $\Sigma_i$.
\foot{
Actually, to be more exact, the operators whose expectation values one
computes this way are not conventional Wilson lines, but receive also
contributions from  the scalar fields.  The
contribution of the scalars is determined, for each $W(P_i)$, by the
choice of a point $R_i\in \S^5$; 
the boundary of $\Sigma_i$ is required to be $C_i\times R_i$.
For details, see
\refs{\newmalda,\reyyee}.  As much as possible, we will suppress the
$R_i$ in the discussion.}

Let us first understand what happens in the low temperature phase.
Here, we seem to have a problem: the $C_i$ are not boundaries of any closed
manifold in $X_1\times \S^5$, so naively all the expectation values
\umpo\ vanish.
Luckily the baryon vertex, as described in \wittenthree, comes to the rescue.
We wrap a D5-brane on $V=Q\times \S^1\times \S^5$, where $Q$ is any point
in $D_4$.  We now meet a phenomenon explained in \wittenthree: because
of the fiveform flux on $\S^5$, a fivebrane can be wrapped on $V$
 only if elementary string worldsheets whose boundaries have
  total winding number $N$ around the
$\S^1$ end on $V$.  We take these boundaries to consist of $N$ circles
$D_i\subset V$ which each wrap around $\S^1$ once.  If $k=N$,
we connect each $C_i$ to $D_i$ by an elementary string worldsheet $\Sigma_i$;
this is possible because $C_i$ and $D_i$ have the same winding number
around the circle.  This configuration with a fivebrane connected to
the circles $C_i$ by elementary string worldsheets gives a nonvanishing
contribution to \umpo\ for $k=N$.  For $k$ any multiple of $N$, one makes 
a similar configuration with several wrapped fivebranes; for $k$ not a multiple
of $N$, there are no such contributions. This 
shows, from the point of view of the low temperature phase, that
the topological symmetry is $\Z_N$.  

Now we consider the high temperature phase, which is described by Type
IIB on $X_2\times \S^5$.  Here we seem to have an opposite puzzle,
which is that the circles $C_i$ are all boundaries in $X_2$; in fact,
since $D_2$ has boundary $\S^1$, $C_i$ is the boundary of $P_i\times
D_2$. Now the expectation value $\langle W(P)\rangle$ of a single
temporal loop on $P\times \S^1$ seems to receive a contribution from a
path integral with an elementary string worldsheet of the form
$\Sigma= P\times
D_2\times R\subset X_2\times \S^5$ ($R$ is a point in $\S^5$ which
actually enters, as mentioned in the footnote above, in the detailed
definition of $W(P)$).  A nonzero expectation value
\eqn\yuppee{\langle W(P)\rangle} would contradict
the existence of the topological symmetry.

This problem was partly resolved in \wittentwo\ by noticing that, with
no cost in energy, we can turn on an NS two-form field $B_{NS} $ 
with $dB_{NS}=0$ and an
arbitrary value of
\eqn\jucog{\alpha=\int_{D_2}B_{NS}.}
Because of the coupling of $B_{NS}$ to the elementary string worldsheet,
the contribution described in the last paragraph to
the expectation value \yuppee\ is proportional to $e^{i\alpha}$.

If there is a symmetry under
\eqn\yoyo{\alpha\to\alpha+\,{\rm constant},}
 then
the expectation value in question will vanish  upon averaging over
$\alpha$ (recall that we are discussing the theory at finite volume,
so we should sum over all possible values of $\alpha$).  
To get the desired result that \yuppee\ vanishes, but \umpo\
is nonzero if $k$ is a multiple of $N$, we want to have a symmetry
not  under arbitrary shifts of $\alpha$, but only under
\eqn\ilcox{\alpha\to \alpha+{2\pi s\over N},\,\,\,s=0,\dots,N-1.}
(To be more precise, 
\ilcox\ will turn out to be a symmetry for any integer $s$,
but if $s$ is divisible by $N$, it is equivalent to a gauge transformation
that is trivial at spatial infinity and so acts as the identity on
 all observables.)
Invariance under arbitrary translations  of $\alpha$ would
correspond to an unwanted $U(1)$ topological symmetry group.

What effects in the theory do not have the invariance \yoyo?
Any term in a low energy effective Lagrangian which is the
integral of a gauge-invariant local density will involve the
$B_{NS}$ field only via $H_{NS}=dB_{NS}$.  So such terms are independent of
$\alpha$.  

A term in the Type IIB low energy effective action which cannot
be written as the integral of a gauge-invariant local density is
\eqn\jubbo{\Delta L=-\int 
B_{NS}\wedge {H_{RR}\over 2\pi}\wedge {G_5\over 2\pi}.}
Here $H_{RR}=dB_{RR}$ is the field strength of the Ramond-Ramond two-form
field $B_{RR}$; $G_5$ is the fiveform field strength. (We have written
the interaction for a spacetime of
Lorentz signature; in Euclidean signature, the minus
sign is replaced by a factor of $i$.) This interaction
is a likely candidate for solving our problem, since \go\ it is intimately
related to the fivebrane wrapping mode which solves the problem in the
low temperature regime.\foot{Actually, because the self-dual fiveform
field is non-Lagrangian, there is a subtlety in the claim
that the Type IIB theory has the interaction $\Delta L$.  One precise
and true statement is that the equations of motion for $B_{NS}$ and $B_{RR}$
contain the terms one would expect by varying $\Delta L$. For our
purposes, one can think of $G_5$ as a fixed background field,
and then the fields $B_{NS}$, $B_{RR}$ can be described by an effective
Lagrangian that contains the term $\Delta L$. Alternatively, we can
write down an effective 5-dimensional Lagrangian for the Type IIB
theory compactified on $\S^5$, and it will contain the terms which
arise from integrating \jubbo\ on $\S^5$.}

Now, in working on the spacetime $X_2\times \S^5$ which is
topologically $\S^3\times D_2\times \S^5$,
we can consider an ``instanton'' in which $H_{RR}$ is pulled back from
the $\S^3$ factor and
\eqn\hoggity{\int_{\S^3}{H_{RR}\over 2\pi}=s,}
with any desired integer $s$. (An important
detail is that the behavior of the metric of $X_2$ near
infinity is such that such a field has finite action despite
the noncompactness of $D_2$.)  We also have 
\eqn\noggity{\int_{\S^5}{G_5\over 2\pi}=N.}
We assume that $B_{NS}$ is pulled back from $D_2$ and obeys \jucog.
Then, we can evaluate
\eqn\poggity{\Delta L = - Ns\alpha.}

The factor $\exp(i\Delta L)$ in the path integral thus equals
$\exp(-iNs\alpha)$.  
The $s=1$ contribution, for example, is invariant precisely under the desired
topological symmetry group \ilcox.

Now, in evaluating \umpo, we get a factor of $e^{i\alpha}$ from each
string world-sheet bounded by one of the $C_i$; this makes a factor of
$\exp({ik\alpha})$ in all.  To get a nonzero result after including
the instanton factor arising from $s$ instantons and integrating over
$\alpha$, we need $\exp(ik\alpha-iNs\alpha)=1$, or $k=Ns$.  Thus, a
nonzero contribution can arise, for some $s$, if and only if $k$ is a
multiple of $N$.  This shows that the topological symmetry is $\Z_N$
also in the high temperature phase.

\subsec{Infinite Volume Limit}

Now we would like to take the infinite volume limit, in the
three-dimensional sense, and replace $\S^3$ by $\R^3$. In the infinite
volume limit the finite temperature theory is always in the high
temperature phase, so we consider Type IIB superstring theory on
$Z=\R^3\times D_2\times \S^5$.

In this limit, one cannot
naturally compute using instantons of the $H_{RR}$ field, because
there are no such localized instantons; a field with
a non-zero value of
$\int_{\R^3}H_{RR}$ will tend to spread out so as to minimize the action.
Nevertheless, since fields with a nonzero integral of $H_{RR}$ gave
the essential contribution in finite volume, it is fairly clear that
treating such fields quantum mechanically must be the key also in infinite 
volume.

The appropriate ideas can be found by a fairly straightforward
adaptation of the analysis of $U(1)$ gauge theory in two dimensions
\ref\coleman{S. Coleman, ``More About The Massive Schwinger Model,''
Annals Phys. {\bf 101} (1976) 239.}.
We write simply $B$ and $H$ for the components of $B_{RR}$ and $H_{RR}$
that are pulled back from the $\R^3$ factor in $Z$.  At $\alpha=0$,
the $H$ field has simply a quadratic Lagrangian of the
form
\eqn\kxon{L_H={1\over {12 e^2}}\int_{{\R^3}}d^3x \,\,
H_{IJK}H^{IJK}}
plus higher order terms (more derivatives or higher powers of $H$)
that will be inessential for our purposes.  Here $I,J,K=1,2,3$,
and $e$ is a constant, independent of $N$, that is obtained by
evaluating the ten-dimensional action for the 
field $H$ that is pulled back from $\R^3$.  (Note 
that $1/e^2$ is finite, despite the noncompactness
of $D_2\times \S^5$.)

To quantize the theory, we split the coordinates as a time coordinate
0 and space coordinates $I=1,2$.  We work in a gauge in which the only
nonzero component of $B$ is $B_{12}$.  A quantum state must be a
gauge-invariant function of $B$ (invariant, that is, under $B\to B+d\Lambda$,
where $d\Lambda$ is a closed two-form with periods that are integer
multiples of $2\pi$).  A basis of states with this property is
furnished by
\eqn\osno{\Psi_r=\exp\left(ir\int_{\R^2}B\right)}
with $r\in \Z$.  The physical interpretation of $\Psi_r$ for $r\not= 0$
is that it describes a state with $H_{012}$ non-zero.
To show this, one notes that if $\Pi(x)=-i\delta/\delta B(x)$ is the canonical 
momentum to $B$, then from the explicit form of $\Psi_r$ we have
\eqn\bosno{\Pi(x)\Psi_r=r\Psi_r}
for all $x$.  On the other hand, since $\Pi(x)=\delta L_H/\delta \del_0 
B(x)$, we get $\Pi=H_{012}/e^2$.  So the state $\Psi_r$ is characterized by
$H_{012}=e^2r$.  Note that this constant value of $H_{012}$ is completely
invariant under the connected component of the three-dimensional
Poincar\'e group, though it is odd under parity.  Since the Hamiltonian
density is
\eqn\hombo{{\cal H}={H_{012}^2\over 2e^2},}
the energy density of the state $\Psi_r$ is
\eqn\gombo{E_r={e^2r^2\over 2}.}

\nref\teitel{J. D. Brown and C. Teitelboim, ``Neutralization Of
The Cosmological Constant By Membrane Creation,'' Nucl. Phys.
{\bf B297} (1988) 787.}

Now we add to the Lagrangian an extra
term
\eqn\vombo{L'_H=-{\theta\over 2\pi}\int d^3x H_{012},}
analogous to the theta term in two dimensional QED.
This does not affect the quantization, the construction of the physical
states $\Psi_r$, or the Hamiltonian \hombo. However, it
adds a constant to $\Pi$, which is now $\Pi=H_{012}/e^2-\theta/2\pi$.
Thus, the three-form field strength is now $H_{012} = e^2 (r + {\theta
\over {2\pi}})$, and the state $\Psi_r$ now has energy
\eqn\bombo{E_r(\theta)= {e^2\left(r+{\theta\over 2\pi}\right)^2\over 2}.} 
(Note that, as in \coleman,
under $\theta\to\theta+2\pi$, there is a  monodromy $E_r\to
E_{r+1}$.  This reflects the fact that $\theta\to\theta+2\pi$ leaves
the theory invariant if accompanied by a gauge transformation that
maps $\Psi_r\to \Psi_{r+1}$.)

Now, going back to our ten-dimensional problem, the interaction \jubbo\
that we used in finite volume reduces on $\R^3$ (after integrating
over $D_2\times \S^5$) to the term \vombo\ with
\eqn\cxon{\theta=-N\alpha.}
The symmetry under $\theta\to\theta+2\pi$ thus becomes the expected
topological symmetry
\eqn\bxon{\alpha\to \alpha+{2\pi\over N}.}
More explicitly, the energy of the state $\Psi_r$ becomes
\eqn\nombo{E_r(\alpha)={e^2\left(r-{N\alpha\over 2\pi}\right)^2\over 2}.}
This formula is clearly invariant
under \bxon\ (if one allows for the possibility
of changing $r$) but not under any additional shifts of $\alpha$.

The effective potential for $\alpha$
is
\eqn\goxon{V(\alpha)=\min_r E_r(\alpha)=\min_r
{e^2\left(r-{N\alpha\over 2\pi}\right)^2\over 2}.}
This function vanishes precisely if $\alpha=2\pi r/N$ for some integer
$r$.  Thus, the topological symmetry is spontaneously broken, and there
are precisely the $N$ vacua required by this symmetry breaking.

Next, let us describe the domain walls between adjacent vacua.
Such a domain wall should look macroscopically like a two-dimensional
surface $\Sigma\subset \R^3$.  Since $r$ jumps by one unit in going
between adjacent vacua, the domain wall has the property
that as one crosses it, the value of $H_{012}$ jumps by one unit.
A $D$-string world-volume has precisely this property. (Note that
in $2+1$ dimensions, a domain wall is in fact a string!)  The
domain wall of the topological symmetry
is thus a $D$-string of world-volume $\Sigma\times P\times Q
\subset \R^3\times D_2\times \S^5$, with $P$ and $Q$ being points in
$D_2$ and $\S^5$, respectively.  In particular, the energy density of
the domain wall is of order $N$ in the 't Hooft large $N$ limit, 
and not of order $N^2$
as one would have expected if the domain wall were a soliton in the large
$N$ effective field theory.  The role of the D-string as a domain wall
is somewhat like what was proposed in \teitel, except that in that discussion
there was no analog of the $\alpha$ field.


\bigskip\noindent{\it Strings Ending On Domain Walls And Duality}

For a final comment on the $SU(N)$ gauge theory, we note the following.  In
$\N=4$ super Yang-Mills theory on $\R^3\times \S^1$,  temporal Wilson
lines have a vacuum expectation value associated with spontaneous breaking
of the topological symmetry.  Meanwhile, spatial Wilson loops
\eqn\hsx{\Tr P\exp\oint_CA,}
with $C$ a large circle in $\R^3$ (at a specified point in $\S^1$)
exhibit an area law, a phenomenon that is equivalent to confinement
in the low energy Yang-Mills theory on $\R^3$.
As discussed in \wittentwo\ (using the framework of \refs{\newmalda,\reyyee}), 
the string or ``flux tube'' associated with
this confinement is simply the elementary Type IIB string.

Since elementary strings can end on $D$-branes, it follows that the
flux tubes associated with confinement can terminate on the domain
walls that separate different vacua.  This is reminiscent of the
behavior found in \ref\ugwitten{E. Witten, ``Branes And The Dynamics
Of QCD,'' Nucl. Phys. {\bf B507} (1997) 658, hep-th/9706109.} 
for chiral domain walls in $\N=1$ super Yang-Mills theory
in four dimensions, on which flux tubes associated with confinement can end.

Now consider the $\tau\to -1/\tau$ duality symmetry of the underlying
Type IIB theory.  In the $\N=4$ super Yang-Mills theory on $\R^3\times \S^1$,
this symmetry exchanges Wilson loops with 't Hooft loops.
Since this symmetry also exchanges fundamental strings with $D$-strings,
it exchanges the flux tubes associated with confinement with the domain
walls associated with spontaneous breaking of the topological symmetry.

In fact, the existence of an $SL(2,\Z)$ duality symmetry in this case
suggests a more symmetric view of the different strings in the
theory. In addition to the angular variable $\alpha$ which we
associated with the $\Z_N$ topological symmetry, there is a similar
angular variable $\beta = \int_{D_2} B_{RR}$, which may be associated
with a magnetic counterpart of the $\Z_N$ symmetry discussed
above. Arguments similar to the ones above show that there is
a ``magnetic'' topological symmetry
 $\beta\to\beta+2\pi / N$, and that the 't Hooft line
may be viewed as an order parameter for the breaking of this $\Z_N$
symmetry. The ``magnetic'' $\Z_N$ symmetry is spontaneously broken in the
finite temperature theory (at infinite volume), and the fundamental
string worldsheets serve as domain walls for the magnetic $\Z_N$
topological symmetry.

\nref\dufflupope{M. J. Duff, H. Lu and C. N. Pope, ``$AdS_5\times \S^5$
Untwisted,'' hep-th/9803061.}
\nref\asy{O. Aharony, J. Sonnenschein and S. Yankielowicz,
``Interactions of Strings and D-branes from M Theory,''
Nucl. Phys. {\bf B474} (1996) 309, hep-th/9603009.}

Altogether, then, there is in
fact a $\Z_N\times \Z_N$ topological symmetry.  The finite
temperature theory has $N^2$ different vacua, corresponding to the
possible values of $\alpha$ and $\beta$. The theory includes $(p,q)$
strings (for relatively prime $p$ and $q$) that serve as domain walls;
upon crossing such a domain wall, $\alpha$ changes by ${2\pi q}\over
N$ and $\beta$ changes by ${2\pi p}\over N$. A $(p_1,q_1)$ domain wall
and a $(p_2,q_2)$ domain wall can merge into a $(p_1+p_2,q_1+q_2)$
domain wall; in the M theory description of the type IIB string theory
(which was described for this type of vacua in \dufflupope) such
string intersections become continuous membrane surfaces \asy.

\newsec{``Topological Symmetry'' Of The $A_{N-1}$ $(0,2)$ Theory In
Six Dimensions}

In six dimensions, there is a family of superconformal theories
with $(0,2)$ supersymmetry and an $A-D-E$ classification.  The $A_{N-1}$
theory is believed \malda\ to be equivalent to $M$-theory on $AdS_7\times
\S^4$, with $N$ units of four-form flux on $\S^4$.  In other words,
if $C$ is the three-form potential and $G=dC$, then to get the 
$A_{N-1}$ theory, one wants
\eqn\iklop{\int_{\S^4}{G\over 2\pi}=N.}

If the $A_{N-1}$ theory is compactified on a circle, we get a five-dimensional
theory with $SU(N)$ gauge group at low energies.  Upon further compactification
on a second circle, a $\Z_N$ topological symmetry  arises which is
associated with the center of this $SU(N)$ gauge theory.  So, 
 the $A_{N-1}$ theory compactified on $\S^1\times \S^1=\T^2$
should have
 a $\Z_N$ global symmetry.  Here we will use the AdS-CFT correspondence
to demonstrate a stronger result: the $(0,2)$ theory actually acquires
a $\Z_N$ global symmetry if it is compactified on {\it any} Riemann
surface $F$.  This statement will hopefully serve as a clue to the
nature of the $(0,2)$ theory.

We consider the $(0,2)$ theory on $\S^4\times F$ (or, taking the volume
of $\S^4$ to
infinity, on $\R^4\times F$).  According
to the general AdS-CFT correspondence, to study the $(0,2)$ theory on
$\S^4\times F$, we must find a seven-dimensional Einstein manifold
$X$, of negative curvature, with $\S^4\times F$ at ``conformal infinity.''
Then we consider $M$-theory on $X\times Y$, where $Y=\S^4$,
 and sum over the contributions
for different choices of $X$.\foot{Several
possible generalizations may be necessary.  For example,
$X$ might contain branes or other singularities,
or it may be incorrect in general to assume that the $M$-theory spacetime
is a product $X\times \S^4$.  Nevertheless, the case stated in the text
seems illustrative.}  (The reason for denoting the second factor of $X\times
\S^4$ as $Y$ is to avoid confusion
 with the $\S^4$ factor in the spacetime $\S^4\times F$ of the
$(0,2)$ model.) For large $N$ this sum will presumably be dominated
for generic values of the parameters by
a single manifold, as in the previous section.

Rather than Wilson lines, the $(0,2)$ theory has two-surface
observables. In the free $(0,2)$ theory with a single tensor multiplet
including a 2-form field $B$ (with self-dual field strength) 
these observables are of the form $\exp i\int_{\Sigma} B$.  But the
generalization of this expression to the interacting $A_{N-1}$ theory
is unknown \foot{The generalization of loop equations to this type of
surface observables was discussed in \ref\ganor{O. J. Ganor, ``Six
Dimensional Tensionless Strings in the Large $N$ Limit,'' Nucl. Phys.
{\bf B489} (1997) 95, hep-th/9605201.}.}. In the context of the
AdS/CFT correspondence,
the observable associated with a closed oriented
two-surface $\Sigma\subset \S^4\times F$
is computed by summing over configurations in which, in $X\times Y$,
there is a membrane whose worldvolume $W$ has $\Sigma$ for boundary.
For $P$ a point in $\S^4$, we let $\Sigma(P)=P\times F$, and we denote the
observable associated with $\Sigma(P)$ as $W(P)$.  We 
will demonstrate the $\Z_N$ global symmetry by showing that expectation values
\eqn\ucucn{\langle W(P_1)W(P_2)\dots W(P_k)\rangle}
vanish unless $k$ is a multiple of $N$.

According to Poincar\'e-Lefschetz duality, the subspace  of $H_*(\S^4\times 
F)$ consisting of homology classes that are boundaries in $X$ is
a ``maximal isotropic subspace.''  It follows that the possible $X$'s are
of two kinds:

(1) $\S^4$ is a boundary in $X$, but $F$ is not.  An
example is $X=D_5\times F$, where the boundary of $D_5$ is $\S^4$.

(2) $F$ is a boundary in $X$, but $\S^4$ is not.  An example is
$X=\S^4\times D_3$, where the boundary of $D_3$ is $F$.

We want to show that every $X$ has the property that $M$-theory on
$X\times Y$ contributes to \ucucn\ only if $k$ is a multiple of $N$.

For $X$ of type (1), the surfaces $\Sigma(P_i)$ are not boundaries
in $X\times Y$, so to get a nonzero contribution to \ucucn, we need
a ``baryon vertex'' on which the membranes that begin on $\Sigma(P_i)$ can
end.   

This can be found as follows.
For $X$ of type (1), a fivebrane wrapped on $F\times Y$ is stable.
It serves as a baryon vertex for the following reason.  Let $T$ be the
self-dual threeform field on the fivebrane.  
Let $W$ be the union of all membrane worldvolumes, and -- bearing in mind
that membranes can end on fivebranes -- let $\partial W$ be the Riemann
surface in the fivebrane worldvolume that consists of boundaries of membranes.
Since the boundaries of membranes in fivebranes are charged under $T$, one has
\eqn\ilo{dT=G-2\pi\delta(\partial W).}
Because of
\iklop, this equation can be solved for  $T$ precisely if  $\partial W$
is homologous to $N$ copies of  $F$.  Thus, the fivebrane on $F\times Y$
contributes to \ucucn\ precisely if $k=N$.  By taking several such fivebranes,
one gets contributions for $k$ any multiple of $N$.

For an $X$ of type (2), 
we have instead that the individual $\Sigma(P_i)$ are boundaries in $X$,
so at first sight it appears that $\langle W(P_i)\rangle$ is nonzero
for each $i$.  However, 
for $X$ of type (2), $F$ is the boundary of a three-cycle $B$ in $X$.
At no cost in action, one can turn on a $C$
field on $X\times Y$  with nonzero
\eqn\hicv{\alpha=\int_{B} C}
but $G=0$.  The matrix element \ucucn\ receives its contribution
from membranes wrapped on a cycle such as $B$, so
the coupling of $C$ to membranes gives an $\alpha$-dependent
factor $e^{ik\alpha}$ to this matrix element.
If this were the only $\alpha$-dependent factor, then integration over
$\alpha$ would cause the matrix element to vanish for all $k$.

To get a nonzero result for suitable $k$, 
we proceed  just as we did in section 2
for the case of the $\N=4$ theory  in  four  dimensions. 
On $\S^4\times F$, we pick a four-form $G$ with
\eqn\umbo{\int_{\S^4}{G\over 2\pi}=s,}
for some $s\in \Z$.  Because $X$ is of type (2), $G/2\pi$ extends over $X$
as a closed fourform with integral periods.
By minimizing its action (an important detail is that the action of $G$ is
in fact finite!) subject to the condition \umbo, $G$ can be turned
into a harmonic four-form that represents an instanton in $M$-theory
on $X\times Y$.  Because of the existence of an interaction in $M$-theory
of the form
\eqn\ijsn{-{1\over 6(2\pi)^2}\int C\wedge G \wedge G,}
and using \iklop,
the amplitude for this instanton has a phase factor $e^{-isN\alpha}$.
By choosing $s=k/N$, one cancels the factor $e^{ik\alpha}$ and
gets a nonzero contribution to \ucucn\ whenever
$k$ is a multiple of $N$. Thus, again we find that the $A_{N-1}$
$(0,2)$ theory compactified on a Riemann surface has a $\Z_N$ symmetry.

\bigskip\noindent{\it Behavior For Infinite Volume}

As in section 2, we can similarly study the $(0,2)$ theory on $\R^4\times F$.
For this, one must study $M$-theory on spacetimes $\R^4\times D_3\times Y$
for some $D_3$.  The key point is to study the dynamics of the field
$G_{0123}$ on $\R^4$.  An analysis as in section 2, using the
interaction term \ijsn, gives an effective
potential $V(\alpha)$ of the same form as in \goxon.  In particular,
the $\Z_N$ symmetry is spontaneously broken; there are $N$ vacua,
at $\alpha=2\pi r/N$, $r=0,1,\dots, N-1$.  The domain walls between
successive vacua are simply the  membranes of $M$-theory.  

One can also consider the observable $W(\Sigma)$ for $\Sigma$
a surface in $\R^4$ (times a point in $F$).  This is analogous to
the ``spatial Wilson loop'' of the $\N=4$ theory on $\R^3\times \S^1$.
The expectation value $\langle W(\Sigma)\rangle$ shows a ``volume law'' 
-- it vanishes exponentially with the volume of a minimal three-surface
of boundary $\Sigma$, because of the appearance of a ``flux surface''
(analogous to flux tubes in confining gauge theories) spanning $\Sigma$.
This can be argued just as in \wittentwo\ for the
spatial Wilson loops of the $\N=4$ theory.  The ``flux surfaces''
are $M$-theory membranes, so in this case the flux surfaces associated
with ``confinement'' also double as ``domain walls'' for spontaneous breaking
of the topological symmetry.  These flux surfaces can end on the domain
walls, since there is no trouble in joining two membranes smoothly.
The fact that, unlike in the previous section, we find only one $\Z_N$
factor may be viewed as the generalization to the interacting theories
of the self-duality of the field strength of the 2-form $B$ of the free
$(0,2)$ theory.

\bigskip
The work of O. A. was supported in part by DOE grant
DE-FG02-96ER40559; the work
of E. W.  was supported in part by NSF Grant PHY-9513835.
\listrefs
\end